\documentclass[sigplan,screen,sigconf]{acmart}

\setcopyright{rightsretained}
\acmDOI{}
\acmISBN{}
\acmConference[LATTE '21]{1st Workshop on Languages, Tools, and Techniques for Accelerator Design}{April 15, 2021}{Virtual, Earth}

\usepackage[utf8]{inputenc}
\usepackage{verbatim}
\usepackage{array}
\newcolumntype{x}[1]{>{\centering\let\newline\\\arraybackslash\hspace{0pt}}p{#1}}

\usepackage[nolist,nohyperlinks]{acronym}
\usepackage{tablefootnote}

\title{Building Beyond HLS: Graph Analysis and Others}

\author{Pedro Filipe Silva}
\affiliation{%
  \institution{Faculty of Engineering, University of Porto}
  \streetaddress{Rua Dr. Roberto Frias s/n}
  \city{Porto}
  \country{Portugal}}
\email{pedro.filipe.silva@fe.up.pt}

\author{João Bispo}
\affiliation{%
  \institution{INESC-TEC and Faculty of Engineering, University of Porto}
  \streetaddress{Rua Dr. Roberto Frias s/n}
  \city{Porto}
  \country{Portugal}}
\email{jbispo@fe.up.pt}

\author{Nuno Paulino}
\affiliation{%
  \institution{INESC-TEC and Faculty of Engineering, University of Porto}
  \streetaddress{Rua Dr. Roberto Frias s/n}
  \city{Porto}
  \country{Portugal}}
\email{nuno.m.paulino@inesctec.pt}

\date{February 2021}

\begin{acronym}
\acro{HLS}{High-Level Synthesis}
\acro{EE}{Electronics Engineering}
\acro{HPC}{High-Performance Computing}
\acro{HBM}{High-Bandwidth Memory}
\acro{ASIC}{Application-Specific Integrated Circuit}
\acro{RTL}{Register-Transfer Level}
\end{acronym}

\begin{abstract}
High-Level Synthesis has introduced reconfigurable logic to a new world -- that of software development. The newest wave of HLS tools has been successful, and the future looks bright. But is HLS the end-all-be-all to FPGA acceleration? Is it enough to allow non-experts to program FPGAs successfully, even when dealing with troublesome data structures and complex control flows -- such as those often encountered in graph algorithms? We take a look at the panorama of adoption of HLS by the software community, focusing on graph analysis in particular in order to generalise to \textit{FPGA-unfriendly} problems. We argue for the existence of shortcomings in current HLS development flows which hinder adoption, and present our perspective on the path forward, including how these issues may be remedied via higher-level tooling.
\end{abstract}

\begin{document}

\maketitle

\section{Introduction}
\label{sec:intro}

The \ac{HLS} approach to FPGA hardware synthesis has quite a storied past. From the 1970s-80s, generations of HLS (or behavioural synthesis, as it was once known) have come and gone \cite{martin_high-level_2009} -- but the current one has shown staying power. Current \ac{HLS} tools have reached new heights in academic development, sophistication, and commercial success \cite{nane_survey_2016}. 

As a result, these tools are now marketed to pure software developers: by many accounts, \ac{HLS} is ready to break into the software world \cite{lahti_are_2019, lant_toward_2020}. The opposing viewpoint persists, however: that implementing FPGA accelerators with current HLS tools still requires significant hardware development knowledge.

We present a somewhat more nuanced viewpoint: \textit{Assuming the existence of higher-level tools such as libraries}, it is perfectly possible for a pure software engineer to implement an FPGA accelerator. Take a relatively mature product such as the the Vitis Vision Library \cite{xilinx_vitis_2021} -- browsing through examples shows the software abstraction mostly holds tight (if one does not descend to the internals). But this begs a question: what happens when there are no viable high-level tools?

\section{What's Missing? (A ``Case Study'' on Graph Analysis)}

Graphs are a common sight in a myriad of scientific and engineering problems and models (\textit{e.g.}, traffic navigation, electrical network analysis) \cite{foulds_graph_1992}, making research on graph algorithms -- and their acceleration -- very active. Notably, graph analysis tends to be both slow and very common in performance-sensitive applications (as is traffic navigation), meaning acceleration is all the more relevant.

FPGA graph accelerators are notoriously difficult. Low global memory bandwidths and low amounts of fast local memory do not mesh with the high memory dependence (often including random access requirements) of graph analytic algorithms. As a result, the combination remains somewhat unexplored \cite{besta_graph_2019}.

Table \ref{tab:fpga-graph-work} shows work on FPGA graph algorithms from 2010 onward, grouped by type\footnote{Results for the \textit{RTL} line were sourced from \cite{besta_graph_2019}, and only cover up to the year 2019.}\footnote{\textit{NPDU} means ``number of works which poorly define programmer usability'' -- we reference this below.}. Note that we define \emph{framework} as a work aiming to provide a high-level generalised tool able to solve several distinct problems, rather than focusing on a specific problem (\textit{e.g.}, shortest path computation).

\ac{RTL} work overwhelms HLS work in number of occurrences. This may indicate that HLS is not yet considered fit for the task, or that \ac{RTL} development methods, versus HLS, still carry significant momentum in the FPGA community. However, recent literature exists arguing that HLS is indeed ready (with some caveats) \cite{lahti_are_2019}, and some novel HLS work claims better results when compared to \ac{RTL} frameworks \cite{chen_thundergp_2021}, so we tend towards the latter option.

\begin{table}[tb]
    \centering
    \caption{FPGA Graph Analytic Work}
    \scriptsize
    \begin{tabular}{x{1.25cm} c c x{4cm}}
    \toprule
    Type & Number & NPDU & Problems \\
    \midrule
    HLS & 4 & N/A & BFS*3, SpMV*2 \\ \midrule
    HLS Framework & 3 & 0 & CCent, PR*2, BFS*2, SpMV*3, SSSP*2, WCC*2, MST, VC \\ \midrule
    \ac{RTL} & 15 & N/A & PR, BFS*8, SSSP*3, APSP, MM, GC \\ \midrule
    \ac{RTL} Framework & 12 & 7 & PR*7, BFS*7, SpMV*2, SSSP*4,  WCC*5, CC, MST, TRW-S, CNN, VC \\
    \bottomrule
    \end{tabular}
    \label{tab:fpga-graph-work}
    
\end{table}

An analysis of the works compiled in Table \ref{tab:fpga-graph-work} brings us to the following observations:

\paragraph{Patterns}
In general, most works on FPGA graph acceleration possess at least one of the following attributes:
\begin{enumerate}
    \item Tackling a single problem.
    \item Not having available/runnable code.
    \item Being either \ac{RTL}-based or unclear about implementation.
\end{enumerate}

\paragraph{The Limits of \ac{RTL}}
When analysing non-HLS literature in particular, a key takeaway is that the majority of works \textit{does} focus on a single problem, and this tends to be a variation of the shortest-path problem. Are pure \ac{RTL} designs hitting an abstraction limit, where other kinds of computations are very difficult to model? Note that several frameworks, especially \ac{RTL} ones, fail to clearly define how a programmer might use them. This may indicate that usage of such frameworks still requires very specialised knowledge.

\paragraph{A Hard Problem}
Much of the literature (including HLS work) is written mostly from a hardware engineer's perspective. This is likely indicative of the special status of graph analysis in the FPGA world: it is an \textit{FPGA-unfriendly}\footnote{Less amenable to implementation than, for instance, streaming-type algorithms. Note that we do not argue that efficient implementation is impossible or nonviable; only that it is \textit{harder}.} problem, still requiring a fair amount of hardware knowledge (and time investment) to tackle, even with HLS. Due to this, adoption of software paradigms, as has occurred in other areas, is slowed. We base our argument for higher-level tooling in this fact.

\paragraph{Symptoms}
All factors appear to be symptoms of a problem: \ac{RTL} cannot fully tackle graph analysis, and HLS isn't quite there yet.

\section{Why are Software Developers Interested in FPGAs?}

The answer is, of course, HLS \cite{lant_toward_2020, cong_high-level_2011}. But why are they interested in using \textit{FPGA boards and expansion cards}?

\paragraph{The Good}
In the HPC community, factors of interest in FPGA technology include:
\begin{enumerate}
    \item Potential time or energy efficiency increase \cite{brown_weighing_2020, muslim_efficient_2017, weller_energy_2017, lant_toward_2020}.
    \item Interest in or perceived architectural advantages \textit{vs.} traditional processing systems \cite{brown_weighing_2020, lant_toward_2020}.
    \item Interest in reconfigurable heterogeneous computing systems \cite{muslim_efficient_2017, weller_energy_2017}.
\end{enumerate}

\paragraph{The Bad}
Conversely, referenced issues include:

\begin{enumerate}
    \item \label{bad-dif} Difficulty in implementation \cite{brown_weighing_2020} \textit{vs.} GPU systems \cite{muslim_efficient_2017}.
    \item \label{bad-port} Lack of portability between FPGA vendors \cite{weller_energy_2017}.
    \item \label{bad-opt} Optimisation challenges \cite{weller_energy_2017}.
    \item \label{bad-qor} Reduced quality-of-results, in some instances, \textit{vs.} \ac{RTL} tools \cite{lahti_are_2019}.
\end{enumerate}

\paragraph{The Lovely}
However, a fact stands out: \textit{all issues could benefit from higher-level tooling}. Raising the level of abstraction is, as history shows, the most direct method of resolving issues \ref{bad-dif} and \ref{bad-port}. As for issues \ref{bad-opt} and \ref{bad-qor}, modern optimising compilers give a hint: in many, or most, scenarios, programming in a high level (systems) language results in more efficient program code -- an effect more visible the more complex the application or target hardware become: precisely where we are heading with FPGA acceleration.

\section{Discussion and Conclusion}
\label{ch:conc}

The time is right for redoubling research into FPGA acceleration, especially in new libraries, abstractions, and toolsets. FPGA usage is increasing, and recent technologies, such as \ac{HBM}, may come to increase popularity even further.

The state of the art in FPGA \textit{graph} acceleration is also shifting: where single-issue, low-level, \ac{RTL} implementations dominated, now appear high-level frameworks, and, most interestingly, \textit{HLS-based} frameworks (we ourselves are evaluating such a framework \cite{chen_thundergp_2021} to implement graph centrality metrics for traffic navigation).

We have used graph analysis to generalise to unfriendly applications, but some of the issues we've raised may very well be applicable to \emph{all} domains of FPGA acceleration. The remainder of this section may be read in either light.

\subsection{Parallels}

Abstraction is a key feature of hardware \cite{coussy_introduction_2009} and software \cite{wirth_brief_2008} engineering alike. Increased abstraction levels are a necessity as system complexity increases. Not unlike the progression, in software engineering, from machine code to assembly to high-level languages\footnote{Although an argument could be made that C presents a faulty/leaky abstraction, due to its closeness to the underlying hardware \cite{wirth_brief_2008, edwards_challenges_2006}. Perhaps perfect abstractions are unattainable, or even undesirable?}, hardware synthesis has evolved from manual design to logic synthesis, and, more recently, high-level synthesis. Could the two parallel paths actually converge at a distant point to a perfectly agnostic behavioural description language?

In 2009, Martin and Smith \cite{martin_high-level_2009} divided the history of \ac{HLS} into three distinct generations, with an upcoming fourth hinging on conquering control flow. This has been done (to what degree, however, is debatable). So will the \emph{fifth} generation focus on raising the abstraction level further, perhaps via higher-level tooling, to conquer unfriendly applications?

\subsection{So, \textit{What's Missing?}}

The way forward for FPGAs goes through HLS -- this is a reoccurring sentiment \cite{lahti_are_2019, lant_toward_2020, pelcat_design_2016}. In fact, \ac{RTL} may be hitting an abstraction wall. Perhaps it will come to be viewed as equivalent to assembly for hardware design \cite{lahti_are_2019}. But research cannot stop at pure HLS.

While HLS has come to significantly raise the abstraction level\footnote{Also not without fault -- as a review of any highly optimised pure HLS quickly indicates (we reference this in Section \ref{sec:intro}).}, this is not sufficient for unfriendly problems such as graph analysis: hardware knowledge remains a necessity in these instances \cite{chen_thundergp_2021, 203245, baptista_raising_2020}. Thus, we raise the following questions:

\begin{itemize}
    \item Are FPGA HLS accelerators currently competitive for graph analysis, or will they become competitive in the near future? If so, how?
    \item Are HLS libraries and toolsets for FPGA acceleration, especially in unfriendly applications, mature enough for use in production? If not, when will they be?
    \item How will increasing support for \ac{HBM} affect the adoption of FPGAs as graph accelerators for unfriendly applications such as graph analysis?
    \item What other promising technologies (\textit{e.g.}, multi-FPGA systems, overlays, dynamic partial reconfiguration, or runtime binary translation) show potential to increase said adoption?
    \item What impact will these technologies have on the potential of FPGAs as general-purpose hardware accelerators?
\end{itemize}

We argue that the continued development and promotion of tools such as frameworks and libraries is \textit{necessary} in order to move the burden of specialised knowledge away from the domain expert. As such, we hold that the future of FPGA acceleration, especially in remnant markets, will heavily depend not only on advances in compiler technology, but also on investment into high-level tooling.

\bibliography{diss}

\end{document}